# Inventions on Drag and Drop in GUI
## A TRIZ based analysis

**Umakant Mishra**

Bangalore, India

http://umakantm.blogspot.com

**Contents**

## 1. Introduction

Drag and drop operation is one of the key capabilities of any Graphical User Interface. The user can do quite complex operations simply by visually dragging and dropping objects from one location to another. It saves user from remembering and typing a lot of commands.

The result of a drag and drop operation may vary depending the type of source object and type of destination object. For example dragging a file and dropping on a folder may copy or move the file to the destination folder, dropping that file to a remote ftp location may upload that file using internet, dropping that file on a printer icon may print that file, dropping that file on the trash can may delete that file, and dropping that file on an executable may play or open or compute or manipulate that file.

Thus a drag and drop operation although prima facie seems to be a simple operation, it can become extremely complicated depending on the type of objects dragged and the type of destination objects selected for dropping. The difficulties of a drag and drop operation may be summarized as follows.

- ⇒ Dragging and dropping tiny objects may require precise mouse movements, which requires a lot of practice. Dragging an undesirable object by slip of finger may yield undesirable result. Similarly, if the finger is slipped form the mouse while dragging, the object may be dropped elsewhere thereby causing an undesirable result.

- ⇒ The drag and drop operation requires both the source window/ container and destination window / container to be visible on the screen. If the destination container / window is not visible on the screen, it is either not possible or requires complex operations to drop the object on the destination container.

- ⇒ Scrolling the screen while dragging to reach the desired location is a difficult operation as the same pointer devise is to be used for both dragging and scrolling operations simultaneously.

- ⇒ The result of a drag and drop operation can be one of the many permutations and combinations depending on the nature and type of source and destination object. For example, while dragging a file icon from one folder to another folder of the same disk the file is moved, but while dragging the same to a folder of another disk, the file is only copied and not moved.

Thus there are many drawbacks of a conventional drag and drop operation. It is necessary to eliminate these difficulties and improve the scope and power of a drag and drop operation.

## 2. Inventions on drag and drop operations

Many inventors have suggested various methods to get rid of various limitations associated with a conventional drag and drop operation. Let's analyze a few of them in the following pages.

**2.1 Auto-scrolling during a drag and drop operation (US Patent 5611060)**

**Background problem**

During a drag and drop operation, the destination may not be visible in some instances because of the limited size of the display window. In that case it is difficult to drag the object precisely on to the destination place.

If we use the conventional scrollbar mechanism to scroll the window when the mouse indicator is on the scrollbar, it may in many cases cause an undesirable scrolling and lead to user frustration.

## Solution provided by the invention

US Patent 5611060 (invented by Belfiore et al., assigned by Microsoft Corporation, issued in March 1997) provides a scrolling during drag and drop operation where the scroll is determined by the location and speed of the mouse indicator. When the mouse indicator is over a predefined area of the window (say at the border) the system compares the speed of the mouse indicator to a predetermined threshold and scrolls the window if the speed is less than the predetermined threshold.

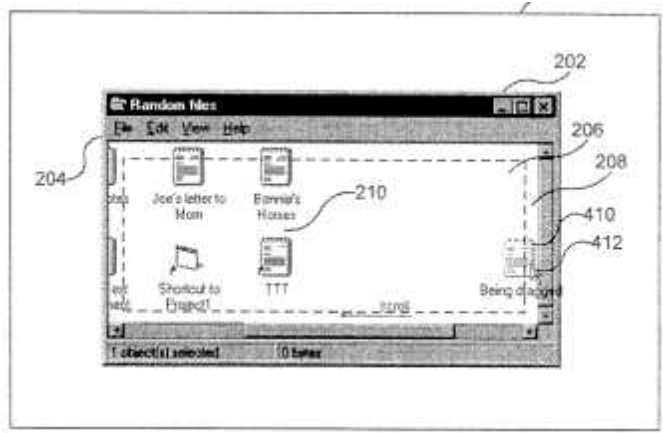

The invention includes a calculation component, a location component and a scrolling component. The calculation component calculates the speed of the mouse pointer while dragging the object. The location component determines whether the screen object is located over a predefined area of the window.

## TRIZ based analysis

The invention provides scrolling of the target window during drag and drop operations (Principle-15: Dynamize).

The system compares the speed of the mouse indicator to a predetermined threshold and scrolls the window if the speed is less than the predetermined threshold (Principle-23: Feedback).

## 2.2 Computer system with graphical user interface including spring-loaded enclosures (US Patent 6061061)

### Background problem

In a drag and drop operation, the user typically places the cursor over an icon to be dragged and moves the cursor into the destination window while the mouse button is depressed. In order to drop the object, the user releases the mouse button when the cursor is over the destination window.

This process is quite cumbersome especially when the destination window is deep in a hierarchy.

**Solution provided by the invention**

Patent 6061061 (invented by Conrad et al., assigned by Apple Computers, issued in May 2000) provides a new behavior which allows the user to open and close enclosure windows while dragging an object. When the user drags an object over a closed enclosure, a temporary window for the closed enclosure is "sprung open" to allow the user to browse inside the enclosure. The user can even open another enclosure contained within the sprung open window, and continue the process to a deep hierarchy until he finds a destination for the drag operation. The user can close the sprung open windows by simply moving the cursor out of the sprung open window.

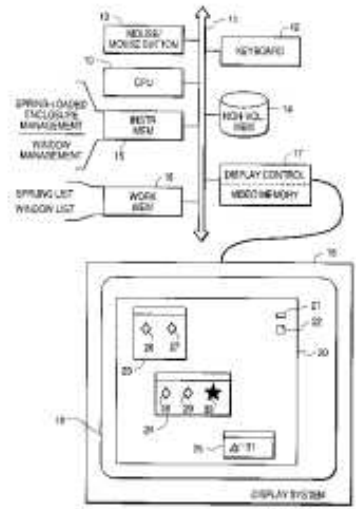

**TRIZ based analysis**

During a dragging operation the folders or containers are automatically opened to show its contents. The containers within the containers are also opened in the hierarchy (Principle-25: Self Service).

The destination containers are opened when the dragging cursor is over the container icon and the containers are closed when the dragging cursor is out (Principle-23: Feedback, Principle-15: Dynamize).

**2.3 Intelligent scrolling (US Patent 6331863)**

**Background problem**

Normally in a drag and drop operation the user drags a source object and drops in a destination object/ window. But in such operation the destination object/ window should be visible at the time of operation. When the data displayed within the window is larger than the display area, some part of the data is "hidden". If a selected item is being dragged to a folder that is "hidden" then the prior art methods do not work, as the destination folder is not visible to be dropped.

There are some prior art solutions to this problem. For example, one may drop the object on a temporary place like desktop, scroll the destination folder to make visible and then drag again from the temporary place to drop on the destination folder. Alternatively one may open the destination folder in a second window and drag the object from the source window to the destination window.

All the above methods of solution need the object to be deposited outside the window. It is desirable to provide for scrolling in a window when items have been selected and are being "dragged" to a folder within the same window.

**Solution provided by the invention**
Patent 6331863 (invented by Meier et al., assignee Apple Computer Inc., issued in Dec 2001) discloses a context sensitive scrolling or intelligent scrolling. The invention allows selecting one or more items in a window and moving the items to a "hidden" destination in the same window. According to the invention, when the user moves the selected items to the edge of the window and pauses for a predefined period of time, the window starts scrolling towards that direction thereby making the "hidden" area visible.

(Picture)

The present invention also works for two windows. When the object is dragged from a first source window to a second destination window, placing the cursor at the edge of the second window can scroll the second window.

**TRIZ based analysis**
The destination folder or window should automatically open up in order to drop the dragged object (Ideal Final Result).

The invention scrolls the window to display the destination location while the user dragging the object (Principle-15: Dynamize).

The window scrolls intelligently based on the position of the dragging cursor thereby making the hidden destination visible (Principle-23: Feedback).

**2.4 Graphical user interface providing consistent behavior for the dragging and dropping of content objects (US Patent 6535230)**

**Background problem**
A drag-and-drop operation is a very useful capability of graphical user interface. The user drags an object, represented by an icon, and drops on another object, represented by another icon. The results of a drag and drop operation depends on several factors, such as type of object dragged and type of destination objects selected. For example, if a text document is dragged and dropped onto a printer icon the operation causes the document to print.

As the types of objects in a GUI environment are plenty the result of a drag and drop operation can be one out of many permutations and combinations, which causes confusion to the user. For example, while dragging a file icon from one folder to another folder of the same disk the file is moved, but while dragging the same to a folder of another disk, the file is only copied and not moved.

It is necessary to ensure consistency of operations when objects are dragged to various destinations in windows and workspaces.

**Solution provided by the invention**

Patent 6535230 (invented by Celik, assigned by Apple Computer Inc., issued in Mar 2003) controls the action of a drag-and-drop operation depending on a single factor, i.e., whether the destination object is a content object or a service object. A content object is an object that has capability to contain another object without manipulating its data. In contrast, a service object is one, which provides a computational result when another object is dropped on it.

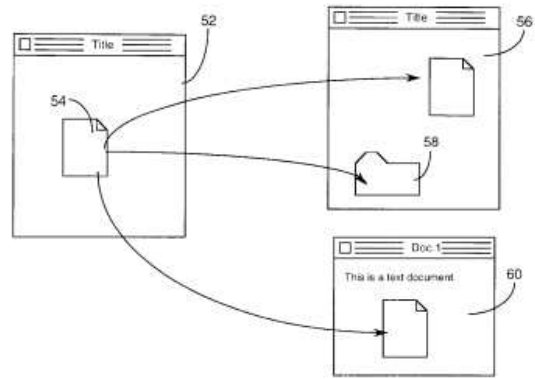

If the destination of a drag-and-drop operation is a content object, the resulting action is to move the dragged object from its original location to the destination. Similarly, if the destination is a service object, the resulting action is to perform the associated service on the dragged object without affecting the source object. If the user desires an operation other than moving, such as copying, then he has to press a specific key on the keyboard during drag-and-drop operation.

**TRIZ based analysis**

The invention standardizes the resulting operation of a drag-and-drop operation in different platforms and applications (Principle-33: Homogeneity).

A normal drag-and-drop operation will do a default action. In order to perform a non-default/ special operation the user will have to press a specific key on the keyboard during the drag-and-drop operation. (Principle-35: Parameter change).

## 3. Summary

Through these pages we found some inventions that have tried to simplify the complicacies of a drag and drop operation. By and large the inventions try to:

- ⇒ Reduce the chances of making mistakes during drag and drop,
- ⇒ Reduce the confusions in drag and drop operations in case of non-standard source or target objects.
- ⇒ Automatic scrolling the screen to display the hidden destinations beyond the visible area of the screen.
- ⇒ Opening up a destination window or container to drop if it is not visible during a dragging operation.
- ⇒ Implementing alternative scroll control mechanism to avoid undesirable scrolling during the operation.
- ⇒ Standardizing operations between different types of objects, such as, data objects, container objects and device objects.